\documentclass[11pt,nofootinbib,preprint]{revtex4}
\usepackage{mathrsfs}
\usepackage{graphicx}
\usepackage{amsmath}
\usepackage{amsfonts}
\usepackage{amssymb}
\usepackage{color}

\usepackage{epsfig}
\usepackage{CJK}
\usepackage{graphicx}
\usepackage{epsfig}
\usepackage{eepic}
\usepackage{bbm}
\usepackage{dcolumn}
\usepackage{bm}
\usepackage{ulem}
\usepackage{slashbox}
\usepackage{multirow}
\usepackage{slashed}

\newcommand{\omits}[1]{}

\def\bc{\begin{center}}

\def\ec{\end{center}}
\def\be{\begin{eqnarray}}
\def\ee{\end{eqnarray}}

\definecolor{dyellow}{rgb}{1.,0.8,.0}
\definecolor{myblue}{rgb}{.1,.1,.7}
\definecolor{dcyan}{rgb}{.0,.6,.6}
\definecolor{cyan}{rgb}{0.4,1.0,1.0}
\definecolor{dmagenta}{rgb}{0.6,0.0,0.6}
\definecolor{brown}{rgb}{0.6,0.2,0.}
\definecolor{darkblue}{rgb}{.0,.0,0.5}
\definecolor{darkred}{rgb}{0.75,0.0,0.0}
\definecolor{orange}{rgb}{1.,.6,.0}
\definecolor{dorange}{rgb}{0.8,.4,.0}
\definecolor{green}{rgb}{0.0,1.0,0.0}
\definecolor{darkgreen}{rgb}{0.0,0.6,0.0}
\definecolor{purple}{rgb}{.4,.0,.4}
\definecolor{lightgrey}{rgb}{0.7, 0.7, 0.7}
\definecolor{grey}{rgb}{0.4, 0.4, 0.4}

\newcommand{\nc}{\newcommand}
\nc{\rnc}{\renewcommand} \nc{\ket}[1]{\left | \, #1 \right \rangle}
\nc{\bra}[1]{\left \langle #1 \, \right |}
\nc{\ua}{\uparrow} \nc{\da}{\downarrow}

\nc{\braket}[2]{\langle\, #1\,|\,#2\,\rangle}
\nc{\half}{\frac{1}{2}}

\nc{\prj}{\mathcal{P}} \nc{\hilb}{\mathcal{H}}
\nc{\pth}{\mathcal{C}} \nc{\inprod}[2]{\braket{#1}{#2}}
\nc{\upket}{\ket{\uparrow}} \nc{\downket}{\ket{\downarrow}}
\nc{\upbra}{\bra{\uparrow}} \nc{\downbra}{\bra{\downarrow}}

\begin{document}


\title{Bit thread, entanglement distillation, and entanglement of purification}

\author{Yi-Yu Lin$^1$} \email{linyy27@mail2.sysu.edu.cn}
\author{Jia-Rui Sun$^{1}$} \email{sunjiarui@mail.sysu.edu.cn}
\author{Yuan Sun$^{1}$} \email{sunyuan6@mail.sysu.edu.cn}

\affiliation{${}^1$School of Physics and Astronomy, Sun Yat-Sen University, Guangzhou 510275, China}


\begin{abstract}
We investigate the relations between bit thread, entanglement distillation and entanglement of purification in the holographic framework. Specifically, we give a bit thread interpretation for the one-shot entanglement distillation (OSED) tensor network, which can be understood as reconstructing the geometric structure of the bulk spacetime from the entanglement information of the boundary quantum system through the ``surface growth scheme''. Moreover, by showing that the holographic entanglement of purification (EoP) process can be regarded as a special case of our ``surface growth scheme'', we naturally obtain the bit thread interpretation of the holographic EoP in our framework, which turns out to be different from the existing interpretations. Since our interpretation for holographic EoP is obtained in a more general and physical framework, the advantage of  our version is that it is more natural, and possibly more reasonable to reflect the real physical entanglement structures.

\end{abstract}


\maketitle
\tableofcontents

\section{Introduction}
In recent years, one of the research directions of holographic principle~\cite{Maldacena:1997re,Gubser:1998bc,Witten:1998qj} is to study the underlying relationship between the entanglement structures in the boundary quantum systems and the geometric structures as well as the dynamics of their dual bulk spacetimes~\cite{VanRaamsdonk:2010pw,Lashkari:2013koa,Faulkner:2013ica,Faulkner:2017tkh,Agon:2020mvu,Bao:2019bib,Sun:2019ycv}. The Ryu-Takayanagi (RT) formula~\cite{Ryu:2006bv,Ryu:2006ef,Hubeny:2007xt} provides the simplest, but probably the most robust clue for this connection. In the holographic setting, in units where $4{G_ N} = 1$ (this convention will always be adopted in this paper), it was shown that, on length scales much larger than the Planck scale, which has been set to 1 in our units, the von Neumann entropy $S(A)$ of a subregion $A$ of the boundary CFT (i.e., the entanglement entropy between $A$ and its complement when the full CFT state is pure) is equal to the area of the bulk minimal extremal surface $\gamma \left( A \right)$ homologous to $A$,
\be\label{rt} S\left( A \right) = {\rm Area}\left( {\gamma \left( A \right)} \right)\ee
Among the subsequent developments, a notable idea is the reformulation of RT formula in the language of bit threads~\cite{Freedman:2016zud,Cui:2018dyq}. Because of the apparent differences between the two descriptions, the study of the relationship between spacetime structures in the bulk and the quantum entanglement in the boundary in the new language may yield some new insights.

Bit threads are unoriented bulk curves which end on the boundary and subject to the rule that the thread density is less than 1 everywhere. According to the so-called max flow-min cut theorem, it can be shown that the maximal flux of bit threads (over all possible bit thread configurations) through a boundary subregion $A$ is equal to the area of the bulk minimal surface homologous to $A$, i.e., the RT surface $\gamma \left( A \right)$. A thread configuration that can achieve this maximal flux is said to $lock$ $A$. Therefore, it is obvious that the RT formula can be expressed in another way, that is, the entropy of a boundary subregion $A$ is equal to the flux of the locking thread configuration passing through $A$~\cite{Freedman:2016zud,Cui:2018dyq},
\be S\left( A \right) = {\rm Flux}_{\rm locking}\left( A \right)\ee
More specifically, in defining the flux, we will define the thread density $\rho$ as the total length of threads contained in a small ball divided by its volume.\footnote{This definition of thread density is the most traditional one, for discussions of other different possible proper definitions of thread density see~\cite{Headrick:2020gyq}.}

In this paper, we will utilize the basic objects in the bit-thread formulation to study the relations between the entanglement structure on the boundary and the geometric structure in the bulk. More specifically, we will reinvestigate the relationship between the entanglement distillation and the emergence of the bulk spacetime from the viewpoint of bit thread. In quantum information theory, it has been shown that for a state ${\left| \varphi  \right\rangle ^{ \otimes m}}$  obtained by the direct product of a large $m$ number of copies of an arbitrary quantum state $\left| \varphi  \right\rangle  \in {\mathcal{H}_A} \otimes {\mathcal{H}_{{A^c}}}$, there exists an operation to approximate it with high fidelity as a state described by a large $n$ number of Bell pairs manifestly encoding the entanglement between $A$ and its complement ${A^c}$, which is called the $entanglement ~distillation$. In particular, it was shown in~\cite{Bao:2018pvs} that the semiclassical holographic limit (i.e., ``large $N$'' and strong coupling) of a single holographic state $\left| \Psi  \right\rangle $ can play the same information-theoretic role as the limit of a large number of identical copies of a single, non-holographic state $\left| \varphi  \right\rangle $ in non-holographic quantum information theory, then the concept of one-shot entanglement distillation (OSED) for a single holographic state and its corresponding OSED tensor network were proposed to describe the mechanism of the AdS/CFT correspondence. Recently, we further investigated and extended the OSED method with the help of the surface/state correspondence~\cite{Miyaji:2015yva,Miyaji:2015fia}, and showed that such kind of tensor networks can be used to describe a ``surface growth picture'', namely, the bulk spacetime geometry can be reconstructed from a continuous growth (propagation) of the bulk minimal surfaces. In other words, the bulk spacetime is emerged by the distillation of the information of the entanglement structures of the boundary quantum system~\cite{Lin:2020thc}.

In the present paper we will present a bit thread interpretation for the OSED tensor network. More precisely, by using the locking theorem put forward and proved by the ``bulk cell decomposition'' method in~\cite{Headrick:2020gyq} recently, we assign a class of bit thread configuration satisfying a series of conditions to match the OSED tensor network, and demonstrate that it describes the finer details of the corresponding OSED tensor network. Furthermore, we will show that the entanglement of purification (EoP) can also be regarded as a special case of the ``surface growth process''. Therefore, our study naturally provides a bit thread interpretation for the EoP, which turns out to be different from the existing interpretations. Since our interpretation version is based on a more general framework, it is more natural, and explicitly shows the physical entanglement structures in the process. After finishing our work, we find an interesting article that discusses the connection between bit threads and the so-called hypergraphs (which also have vertices and edges similar to the tensor networks) is noteworthy~\cite{Bao:2020uku}, although they also discuss non-holographic states therein.

The basic framework of this paper is as follows. In the section \ref{sec2}, we start with reviewing the bit thread locking theorem, and then present our motivation and proposal. In section \ref{sec3} we further clarify the connection between the bit thread and the OSED tensor network. In section \ref{sec4}, we discuss the relation between bit thread and the entanglement of purification. The conclusion and discussion are given in section \ref{sec5}.


\section{Motivation}\label{sec2}
\subsection{A brief review of Bit thread locking theorem}\label{sec2a}
Consider a $d$-dimensional compact Riemannian manifold-with-boundary $M$, for example, it can be a time slice of AdS$_{d + 1}$ spacetime, and then divide its boundary into adjacent non-overlapping subregions ${A_1}, \ldots ,{A_n}$, which is referred to as $elementary~regions$, while a $composite~ region$ is defined as the union of some certain elementary regions. Then, one can express the two existence theorems of locking thread configuration in~\cite{Cui:2018dyq} as follows:

{\bf Theorem 1.} There exists a multiflow that locks all the elementary regions ${A_i}$.

{\bf Theorem 2.} There exists a multiflow that can lock all the elementary regions and any single composite region simultaneously.

Recently, a weak continuum locking theorem was further proved in~\cite{Headrick:2020gyq} by using a ``bulk-cell decomposition'' method (which perhaps should be called ``bulk-cell gluing'' method more appropriately), i.e.,

{\bf Theorem 3.} There exists a multiflow that can lock all the elementary regions and all non-crossing composite regions simultaneously.

Here we are following the terminology from network theory: two boundary regions are said to cross if they partially overlap and do not cover the whole boundary. For example, $AB$ crosses $BC$, but does not cross $A$, $ABC$, or $D$.

\begin{figure}[htbp]     \begin{center}
		\includegraphics[height=9cm,clip]{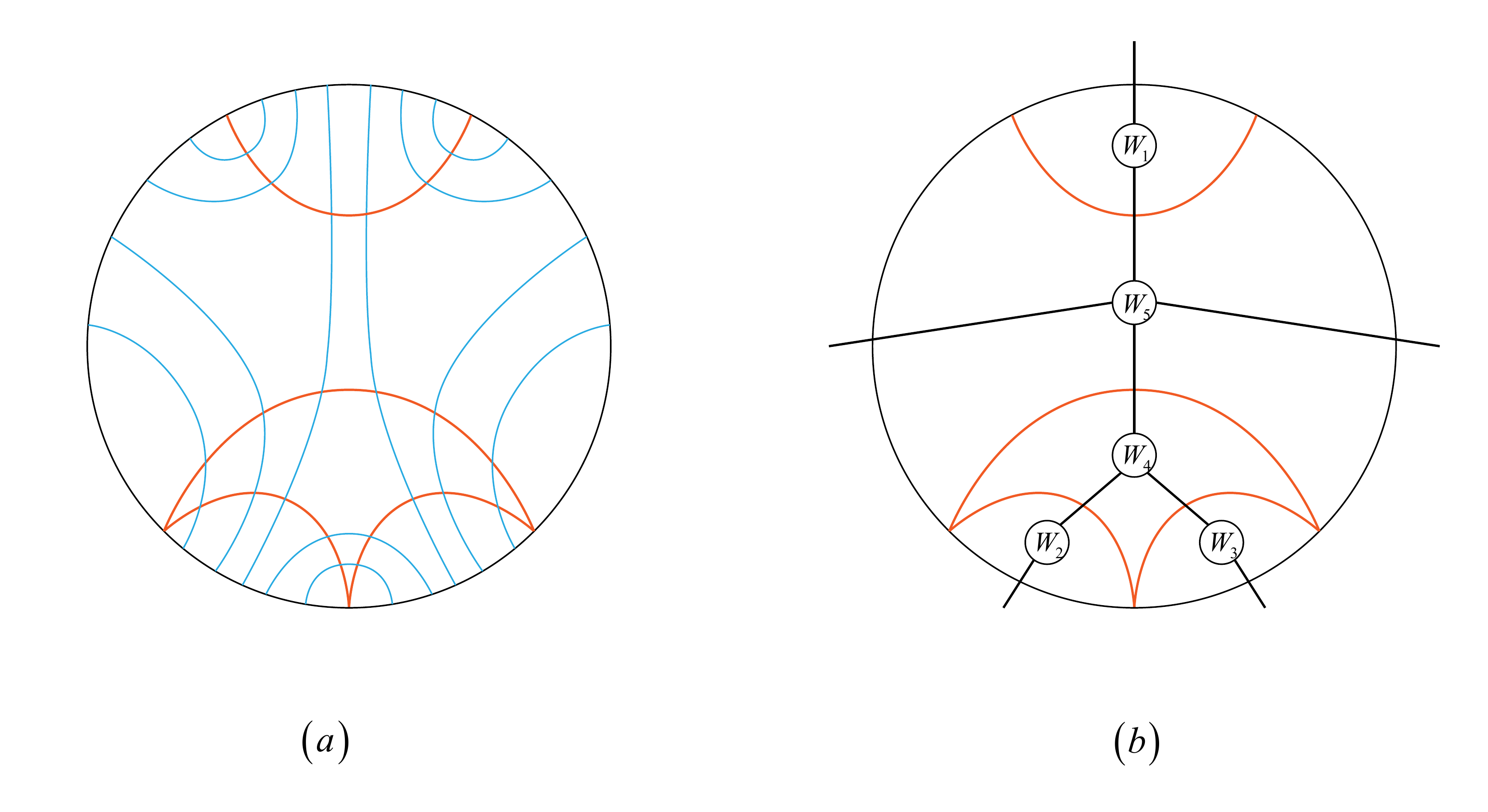}
		\caption{(a) A locking thread configuration constructed by the bulk-cell decomposition. The red lines represent the RT surfaces, and the blue lines represent the bit threads. (b) An OSED tensor network. Here we present the second representation in~\cite{Lin:2020thc}, and leave out the orientation of the bonds.}
		\label{fig00}
	\end{center}	
\end{figure}

As shown in figure~\ref{fig00}(a), the intuitive proof method can be described as follows: first, utilizing a series of RT surfaces according to a certain appropriate order to discretize the bulk spacetime. These small fragments cut by RT surfaces can be called ``cells'', and each cell is bounded by RT surfaces or boundary subregions. Next, regarding each cell as an individual spacetime, then the locking ${\bf Theorem~1}$ states that one can always find a locking thread configuration to lock all boundaries of each cell spacetime simultaneously. Since the boundaries of the cell spacetime are minimal surfaces, the locking thread configuration must satisfy that the threads intersecting these surfaces should be orthogonal to them and saturate the density bound. Finally, be gluing all the locking thread configurations of cell spacetimes together, one can obtain a locking thread configuration of a whole bulk spacetime, and thus prove the existence.

\subsection{Locking thread configuration and OSED tensor network}\label{sec2b}

The bulk cell decomposition method involved in ${\bf Theorem~3}$ is reminiscent of the construction of the OSED tensor network~\cite{Bao:2018pvs, Lin:2020thc}, where one utilizes a series of nonintersecting RT surfaces to discretize the bulk spacetime into cells similarly, see figure~\ref{fig00}(b). A natural and interesting idea is to match the bulk cell decomposition picture of the locking bit thread configuration with the corresponding OSED tensor network. In this way we can provide a bit thread interpretation for OSED tensor network. Furthermore, since the OSED tensor network implies the emergence of bulk spacetime can be regarded as a continuous reorganization of the degree of freedom of boundary system through the process of entanglement distillation~\cite{ Lin:2020thc}, we expect that the bit thread viewpoint might provide more insight into such a emergence picture.

In order to see that this idea is reasonable,
the first thing one can notice immediately is that the appropriate order in the bulk cell decomposition happens to be consistent with the orientation of the corresponding OSED tensor network, see figure~\ref{fig00}. This can be understood as that essentially the ``appropriate order'' implies that the information should always be propagated from the places containing adequate information to ensure the ``reconstruction'' of the next places. The only difference between the two descriptions is that, the carrier carrying the information is thread in the locking thread configuration, while in the OSED tensor network it is the states on the minimal surfaces that carry the information.

\begin{figure}[htbp]     \begin{center}
		\includegraphics[height=12cm,clip]{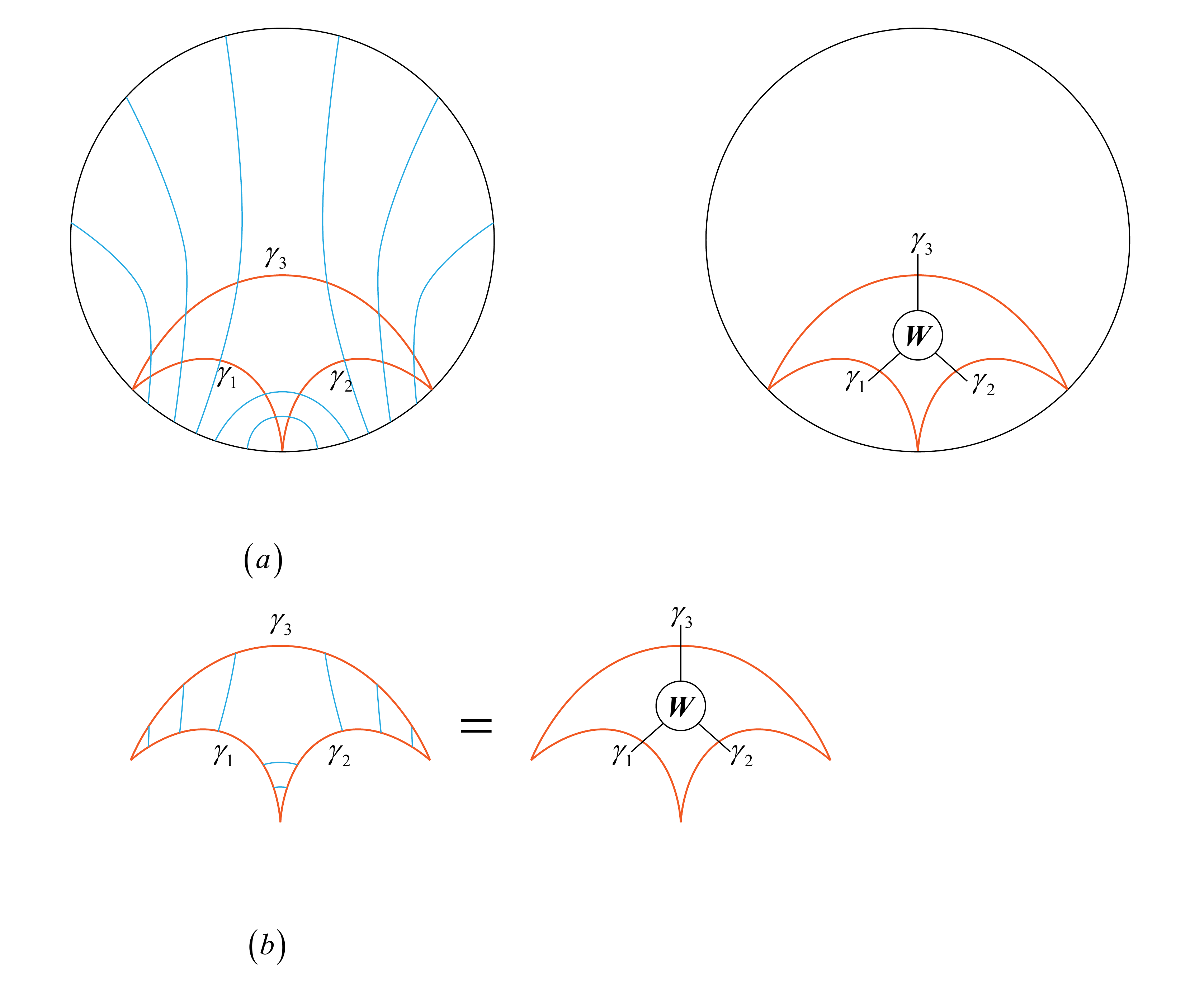}
		\caption{(a) The identification between the locking	thread configuration constructed by the bulk-cell decomposition with its corresponding OSED tensor	network. (b)  The bit thread interpretation of the $W$ tensor in the OSED tensor network.}
		\label{fig1}
	\end{center}	
\end{figure}

By investigating the simplest case of a single cell, we can further demonstrate the reasonability of this equivalence between these two pictures. As shown in figure~\ref{fig1}(a), supposed we identify the locking thread configuration constructed by the bulk cell decomposition with its corresponding OSED tensor network, we will find that this new interpretation from bit thread exactly provides a detailed implementation of the central description in OSED tensor network~\cite{Lin:2020thc}: 

The fact that the von Neumann entropy of minimal surface ${\gamma _3}$ is less than the sum of the entropy of minimal surface ${\gamma _1}$ and that of minimal surface ${\gamma _2}$ actually stems from the entanglement between ${\gamma _1}$ and ${\gamma _2}$ entirely, while there are no internal entanglement within the minimal surfaces per se.

More specifically, if we interpret the bit threads directly connecting the ${\gamma _1}$ and ${\gamma _2}$ in the figure as intuitively representing the entanglement of the two previous minimal surfaces, while interpret the bit threads connecting ${\gamma _1}$ (or ${\gamma _2}$) and ${\gamma _3}$ as representing the information of ${\gamma _3}$ arises from the reorganization of the information of ${\gamma _1}$ and ${\gamma _2}$, then we recover the central description in OSED tensor network. Due to the bit threads directly connecting surface ${\gamma _1}$ and surface ${\gamma _2}$, when tracing the bit threads flowing from the union of ${\gamma _1}$ and ${\gamma _2}$ to ${\gamma _3}$, we find that the flux of threads decreases. Recall that the thread flux through a minimal surface in the locking thread configuration is exactly equal to the von Neumann entropy of this minimal surface itself. Therefore, it is reasonable and natural to endow the OSED tensor network with this bit thread interpretation. In other words, we have obtained the bit thread interpretation of the $W$ tensor in the OSED tensor network.

The essential point is that, on the one hand, we can describe the configuration of bit threads in the language of flow using the divergenceless vector field, therefore, according to Gauss's theorem, the total flux through the closed surface  ${\gamma _1} \cup {\gamma _2} \cup {\gamma _3}$ is equal to zero. However, on the other hand, the quantity really needed for the calculation of entropy is the flux of threads, instead of the flux of vector field. Therefore, when calculating the total thread flux through ${\gamma _2}$, the flux flowing from ${\gamma _1}$ into ${\gamma _2}$ should not cancel out the flux flowing out of ${\gamma _2}$ towards ${\gamma _3}$. One should add their absolute value to obtain the total thread flux through ${\gamma _2}$. In particular, when the geometry of cell has left-right symmetry, the thread fluxes on both sides of ${\gamma _1}$ and ${\gamma _2}$ can be guaranteed to be equal to preserve the symmetry.

However, note that the configuration that can lock the given decomposition is not unique. Moreover, in the original configuration proposed in~\cite{Headrick:2020gyq}, as shown in figure~\ref{fig1}(a), there seems to be a problem of oversimplification. In this kind of configuration, the threads connecting ${\gamma _1}$ and ${\gamma _2}$ are all located at the inner regions of both surfaces, while the threads connecting ${\gamma _1}$ (or ${\gamma _2}$) and ${\gamma _3}$ are all located at the outer regions. This is essentially because in general one tends to require the bit threads satisfying the local parallelity. However, according to the above viewpoint that the locking thread configuration can provide a concrete description of the surface growth process, there is no reason to suppose that there are no bit threads representing the mutual entanglement in the outer regions. Similarly, the information in the inner region should also contribute to the information of surface ${\gamma _3}$.

\section{The locking thread configuration corresponding to OSED tensor network}\label{sec3}

\begin{figure}[htbp]     \begin{center}
		\includegraphics[height=8cm,clip]{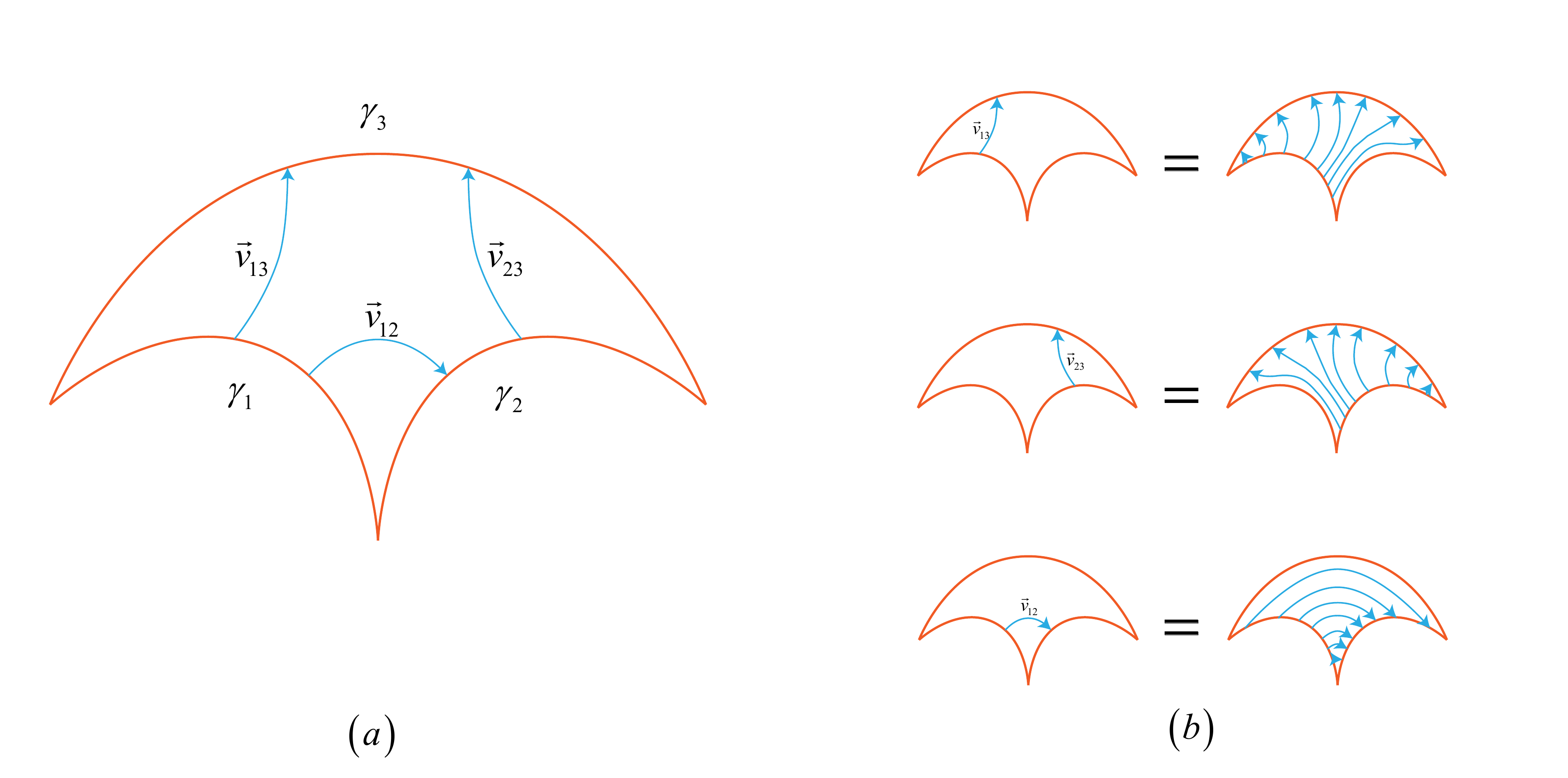}
		\caption{(a) A multiflow $V = \{ {({\vec v_{ij}})_{i < j}}\} $ representing the locking thread configuration. (b) For simplicity, we use a single thread to represent a thread bundle, which is described by a component flow.}
		\label{fig2  }
	\end{center}	
\end{figure}

As we have seen clearly in the previous section, the locking bit thread configuration is very suitable to describe the OSED process, since they are constructed in a similar way and the locking thread configuration captures the key characteristic of the OSED process. However, in order to further sharpen this relationship, in this section we would consider the more general locking thread configuration, in the spirit of bulk cell gluing method presented in~\cite{Headrick:2020gyq}, but presenting more details. 

We will use  a multiflow $V = \{ {({\vec v_{ij}})_{i < j}}\} $ to represent the locking thread configuration (Note that in the present paper we will define ${\vec v_{ij}}$ only with $i < j$ for convenience, which is slightly different from (but equivalent) convention adopted in~\cite{Cui:2018dyq}, where the fields ${\vec v_{ij}}$ were also defined for $i \ge j$, but with the constraint ${{\vec v}_{ji}} =  - {{\vec v}_{ij}}$). Let us still take the simplest case involving only three RT surface as the first example. As shown in figure~\ref{fig2  }, we define three thread bundles represented by three component flows ${\vec v_{12}}$, ${\vec v_{13}}$ and ${\vec v_{23}}$ respectively~\footnote{It should be noted that even in this case, according to the usual definition of multiflow, the thread bundle corresponding to each component flow is still locally parallel. For example, as shown in figure~\ref{fig2  }(b), it seems that the thread bundle corresponding to ${\vec v_{12}}$ characterizes the entanglement between ${\gamma _1}$ and ${\gamma _2}$ in an artificial symmetric pattern. Nevertheless, as will be shown later, actually this kind of particular choice is not artificial, but can be naturally understood as a reflection of the coarse-graining.}. In addition to satisfying the basic condition for multiflow,
\be\label{gauss}\nabla  \cdot {\vec v_{ij}} = 0\ee
\be\label{bypass}{\left. {\hat n \cdot {{\vec v}_{ij}}} \right|_{{\gamma _k}}} = 0\,\,\,\,\,({\rm{for}}\,\,k \ne i,j{\rm{)}}\ee
\be\rho (V) = \sum\limits_{i < j} {\left| {{{\vec v}_{ij}}} \right|}  \le 1\ee
in the locking thread configuration, on the minimal surfaces ${\gamma _i}$
 they must also satisfy
\be{\vec v_{ij}}\,{\rm{is}}\,{\rm{orthogonal~to}}\,{\gamma _i}\,{\rm{or}}\,{\gamma _j}\ee
and
\be{\left. {\rho (V)} \right|_\gamma } = 1.\ee
Defining the flux of a vector field ${\vec v}$ through a region $A$ as
\be\int_A {\vec v}  = \int_A {\sqrt h \hat n \cdot \vec v}, \ee
and for convenience, we denote the value of the flux of the bit threads represented by component flow ${\vec v_{jk}}$ (i.e., the absolute value of the flux of vector field ${\vec v_{jk}}$ per se) through surface ${\gamma _i}$ as
\be F{(i)_{jk}} = \left| {\int_{{\gamma _i}} {{{\vec v}_{jk}}} } \right|,\ee
while denote the total flux of bit threads through surface ${\gamma _i}$ as
\be N\left( i \right) = \sum\limits_{j,k} {F{{\left( i \right)}_{jk}}}. \ee
Using Gauss's theorem for (\ref{gauss}), for the case in figure~\ref{fig2  }, one can obtain
 \be\label{f1} F{\left( 3 \right)_{13}} - F{\left( 1 \right)_{13}} = 0,\ee
 \be\label{f2} F{\left( 3 \right)_{23}} - F{\left( 2 \right)_{23}} = 0,\ee
 \be\label{f3} - F{\left( 1 \right)_{12}} + F{\left( 2 \right)_{12}} = 0,\ee
which, in a more compact form, is
 \be\sum\limits_{i < j} {\int_{{\gamma _k}} {{{\vec v}_{ij}}} }  = 0.\ee
Since the multiflow locks the three minimal surfaces, the entropy $S(i)$ of the minimal surface ${\gamma _i}$ is equal to the flux of bit threads,
\be\label{s3} S(3) = N\left( 3 \right) = F{(3)_{13}} + F{(3)_{23}},\ee
\be\label{s1} S(1) = N\left( 1 \right) = F{(1)_{12}} + F{(1)_{13}},\ee
\be\label{s2} S(2) = N\left( 2 \right) = F{(2)_{23}} + F{(2)_{12}},\ee
in a more compact form, that is
\be S({\gamma _i}) = \sum\limits_{j < k} {\left| {\int_{{\gamma _i}} {{{\vec v}_{jk}}} } \right|} \ee
By substituting eqs.(\ref{f1})(\ref{f2})(\ref{f3}) into eqs.(\ref{s1})(\ref{s2})(\ref{s3}), one can obtain
\be\label{ssadd} S(1) + S(2) - S(3) = 2F{(1)_{12}} > 0,\ee
just as we expect from the strong subadditivity of the entanglement entropy. Eq.(\ref{ssadd}) can also be rewritten as the constraint on the component flow in the locking thread configuration, i.e.,
\be\label{f112} F{(1)_{12}} = \frac{1}{2}\left( {S(1) + S(2) - S(3)} \right) = \frac{1}{2}\left( {S(1) + S(2) - S(12)} \right) \equiv \frac{1}{2}I\left( {1:2} \right),\ee
where $I\left( {i:j} \right)$ is the mutual information between ${\gamma _i}$ and ${\gamma _j}$, and in the second equation, the surface/state correspondence~\cite{Miyaji:2015yva,Miyaji:2015fia} has been used in order to regard the state of the closed surface ${\gamma _1} \cup {\gamma _2} \cup {\gamma _3}$ as a pure state, such that $S\left( 3 \right) = S\left( {12} \right)$.

Similarly, we can obtain the constraints on the value of the other two fluxes,
\be F{(1)_{13}} = \frac{1}{2}I\left( {1:3} \right),\ee
\be F{(2)_{23}} = \frac{1}{2}I\left( {2:3} \right).\ee
In principle, any locking thread configuration satisfying the above constrains is eligible to describe the surface growth process, i.e., the entanglement distillation process. From the equivalence of net result, the locking thread configuration in figure~\ref{fig1} can be viewed as a special choice of our prescription.

\begin{figure}[htbp]     \begin{center}
		\includegraphics[height=10cm,clip]{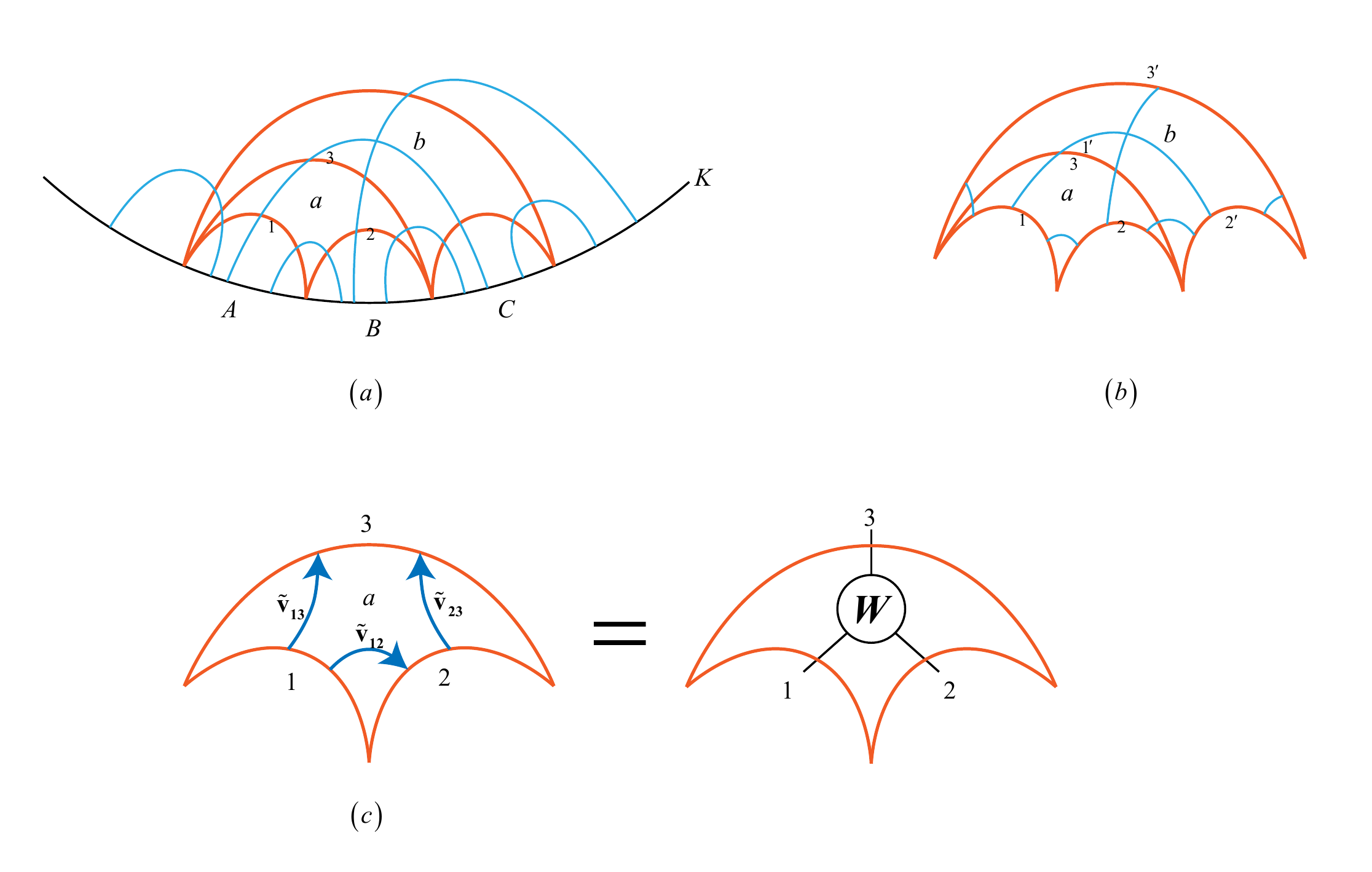}
		\caption{(a) A more general case consisting of elementary cells. For simplicity, we use a single thread to represent a thread bundle. (b) Constructing the whole locking thread configuration by gluing each adjacent elementary cells. (c) The bit thread representation of the fundamental tensor $W$ tensor in the OSED tensor network.}
		\label{fig3  }
	\end{center}	
\end{figure}

\begin{figure}[htbp]     \begin{center}
		\includegraphics[height=8cm,clip]{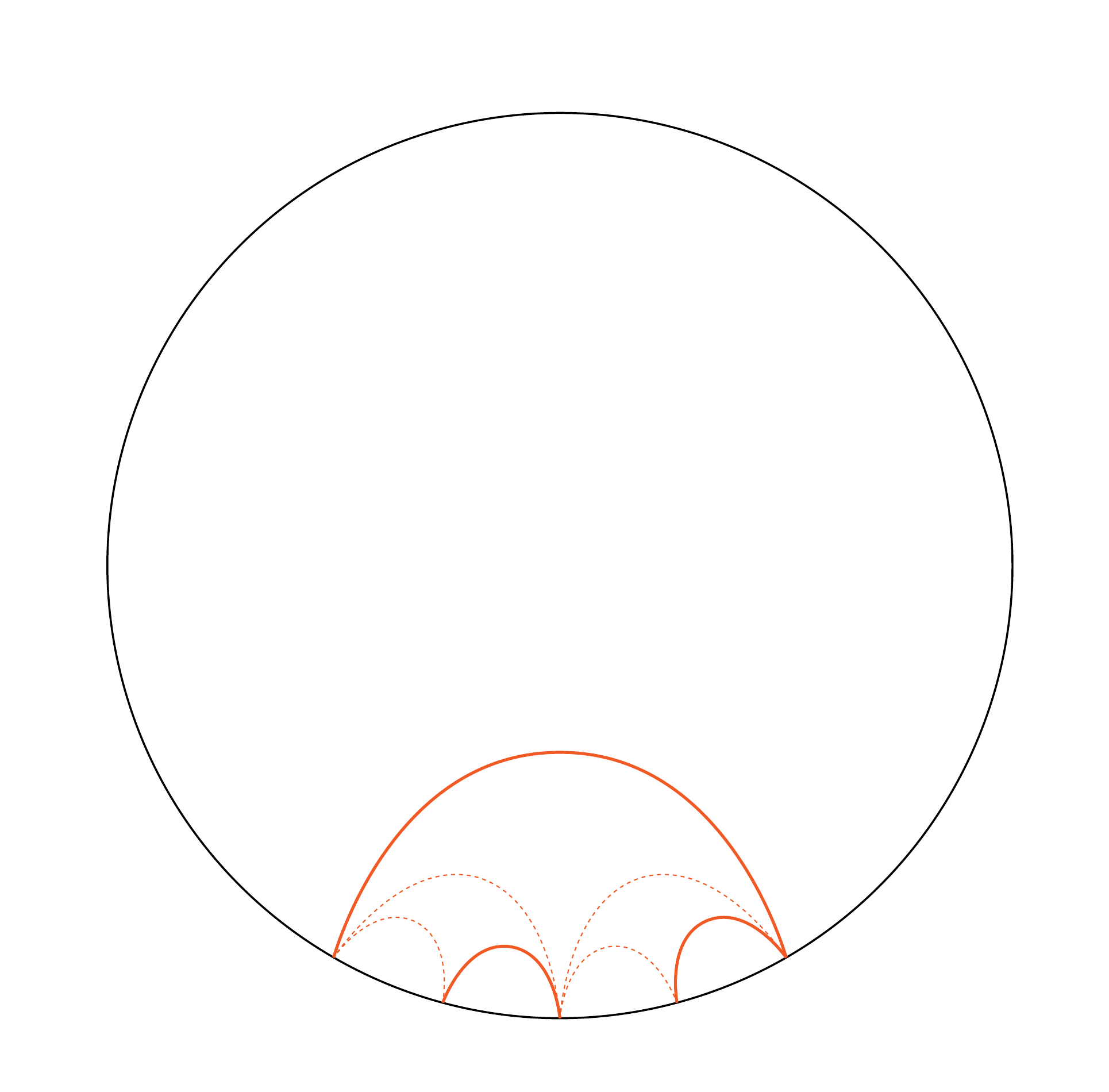}
		\caption{For the more general case, one can always add extra auxiliary RT surfaces to get back to the basic case above.}
		\label{fig4  }
	\end{center}	
\end{figure}
Now let us move on to the more general case involving more than three RT surfaces. The basic idea is the same as that in~\cite{Headrick:2020gyq}, namely, we first construct the locking thread configuration for each small cell spacetime, then glue all these cells together to obtain the locking thread configuration for the whole spacetime. However, we will present more details on the basis of the illuminating picture of~\cite{Headrick:2020gyq}, so as to explore the relation between the locking thread configuration and OSED tensor network appropriately. For convenience and without loss of generality, we assume that all the cells involved in the considered decomposition picture are all surrounded by two adjacent minimal surfaces and a ``new growing'' minimal surface anchored on their outer boundaries~\footnote{Moreover, in this section we will consider the cases only involving RT surfaces, i.e., the minimal surfaces whose anchor points are located on the boundary of the whole spacetime. For more general cases involving bulk minimal surfaces, the spirit of the discussion is the same, but the details of discussion will be more complicated. However, in section~\ref{sec4} we will present an interesting case involving bulk minimal surfaces, which actually represents the entanglement of purification process.}, as shown in figure~\ref{fig3  }(a). We will name this kind of cell as the ``elementary cell'', and in particular denote its three boundaries as ${\gamma _1}$, ${\gamma _2}$ and ${\gamma _3}$ respectively, or more simply, as $1$, $2$, $3$.
Accordingly, we will investigate the OSED tensor networks only containing the $W$ isometry tensors that map two minimal surfaces to one new minimal surface. For the more general case shown in figure~\ref{fig4  }, using the property of the RT surfaces, one can always add extra auxiliary RT surfaces to get back to the basic case we discussed.

Taking the case involving only four elementary regions $A$, $B$, $C$ and their complement $K$ as shown in figure~\ref{fig3  }(a) as an example, our aim is to argue the existence of the locking thread configuration corresponding to this kind of general surface growth picture in the spirit of bulk cell gluing. To describe the locking thread configuration in figure~\ref{fig3  }(a) explicitly, we first denote the elementary regions on the boundary as ${A_i}$, and denote the minimal surfaces involved in the figure as ${\gamma _a}$. Now consider a thread bundle that starts from an elementary region ${A_i}$ and ends up in another elementary region ${A_j}$, and crosses through a series of minimal surfaces, which are collectively denoted as a set ${\left\{ \gamma  \right\}^{\rm{cr}}}$. We can thus denote the component flow describing this thread bundle as ${\vec v_{{A_i}{{\left\{ \gamma  \right\}}^{\rm{cr}}}{A_j}}}$, and denote the set of all the involving flows as a multiflow $\left\{ V \right\}$. Similarly, we adopt the consistent convention as before, i.e., ${\vec v_{{A_i}{{\left\{ \gamma  \right\}}^{\rm{cr}}}{A_j}}}$ is defined only for $i < j$, and its orientation is from ${A_i}$ to ${A_j}$. Then we require ${\vec v_{{A_i}{{\left\{ \gamma  \right\}}^{\rm{cr}}}{A_j}}}$ satisfying
\be\label{req1}\nabla  \cdot {\vec v_{{A_i}{{\left\{ \gamma  \right\}}^{\rm{cr}}}{A_j}}} = 0,\ee
\be\label{req2}\rho \left( V \right) \le 1,\ee
\be\label{req3}{\left. {\hat n \cdot {{\vec v}_{{A_i}{{\left\{ \gamma  \right\}}^{\rm{cr}}}{A_j}}}} \right|_{{\gamma _a} \notin {{\left\{ \gamma  \right\}}^{\rm{cr}}}}} = 0,\ee
\be\label{req4}{\vec v_{{A_i}{{\left\{ \gamma  \right\}}^{\rm{cr}}}{A_j}}} \bot {\gamma _a},\quad{\rm{for}}\quad{\gamma _a} \in {\left\{ \gamma  \right\}^{\rm{cr}}},\ee
\be\label{req5}{\left. {\rho \left( V \right)} \right|_{{\gamma _a} \in {{\left\{ \gamma  \right\}}^{\rm{cr}}}}} = {\sum\limits_{i < j} {\left. {\left| {{{\vec v}_{{A_i}{{\left\{ \gamma  \right\}}^{\rm{cr}}}{A_j}}}} \right|} \right|} _{{\gamma _a} \in {{\left\{ \gamma  \right\}}^{\rm{cr}}}}} = 1.\ee
The first two requirements are following the basic convention of multiflow. Eq.(\ref{req3}) is the extension of eq.(\ref{bypass}), since we want each thread bundle to ``bypass'' the minimal surfaces it does not cross through, this requirement is essentially to ensure that the thread configuration in each cell spacetime can satisfy eq.(\ref{bypass}). The last two requirements are to ensure that this set of multiflow can lock each involved minimal surfaces. In addition, we also require that no thread intersects the same minimal surface twice or more.

Now we can prove the existence of the locking thread configuration described above in the spirit of bulk cell gluing. For this purpose, we consider two adjacent elementary cells, such as the cell $a$ and cell $b$ in figure~\ref{fig3  }, which share a same minimal surface, denoted as $3$ or ${1'}$. Regarding each cell as an individual spacetime, then by locking ${\bf Theorem~1}$, one can always find the locking thread configuration to lock the three boundaries of the cell simultaneously, as long as it satisfies the following conditions
\be\nabla  \cdot \vec v_{ij}^{\rm{m}} = 0,\ee
\be\rho (V) \le 1,\ee
\be{\left. {\hat n \cdot \vec v_{ij}^{\rm{m}}} \right|_{{\gamma _k} \ne {\gamma _i}\,{\rm{or}}\,{\gamma _j}}} = 0,\ee
\be\label{orth}\vec v_{ij}^{\rm{m}} \bot {\gamma _i}\,{\rm{and}}\,{\gamma _j},\ee
\be\label{satu}{\left. {\rho \left( V \right)} \right|_{{\gamma _i}\,{\rm{or}}\,{\gamma _j}}} = {\sum\limits_{{\rm{m}},\,i < j} {\left. {\left| {\vec v_{ij}^{\rm{m}}} \right|} \right|} _{{\gamma _i}\,{\rm{or}}\,{\gamma _j}}} = 1,\ee
where $i$ and $j$ can take $1$, $2$, $3$ in cell $a$, or $1'$, $2'$, $3'$ in cell $b$. The more details presented here than in~\cite{Headrick:2020gyq} is that the bit threads connecting surface $i$ and surface $j$ are allowed to be composed of several independent thread bundles, and we label the different bundles with $m$. The key point is that, in the two adjacent locking processes, according to formulas eq.(\ref{orth}) and eq.(\ref{satu}), on the one hand, in cell $b$, the thread configuration stemming from the shared minimal surface must intersect it orthogonally and saturate the density bound. More specifically, each thread bundle must be orthogonal to the shared surface, and the sum of the fluxes of each bundle through this surface should be equal to the area of this minimal surface itself. On the other hand, similarly, in cell $a$, the thread configuration reaching the shared minimal surface must also intersect it orthogonally and saturate the density bound. Therefore, due to the continuity, one can always assign the threads passing through the shared surface into $m$ independent thread bundles according to the same prescription, and then gluing each thread bundle in both sides respectively. Finally, when gluing all the cells containing the set of minimal surfaces ${\left\{ \gamma  \right\}^{{\rm{cr}}}}$ together, we can obtain the component flow ${\vec v_{{A_i}{{\left\{ \gamma  \right\}}^{{\rm{cr}}}}{A_j}}}$ describing the corresponding thread bundle. We thus complete the proof of the existence of the multiflow proposed above.

It should be noted that the locking thread configuration allowed in each cell spacetime is not unique. This is because the locking condition above only restricts the behavior of the bit threads on the minimal surfaces, while there still exists freedom for the allowed thread configuration in the interior of the cell. In addition, there is also freedom in assigning the bit threads connecting surface $i$ and surface $j$ into several thread bundles represented by $\vec v_{ij}^{\rm m}$. In particular, the illuminating picture presented in~\cite{Headrick:2020gyq} is a special case of our general construction. They assigned the ${\vec v_{ij}}$ into a set of allowed $\vec v_{ij}^{\rm{m}}$ such that the final total thread configuration is locally parallel. In this situation, the multiflow can be directly transferred into a single flow to describe the thread configuration equivalently. However, the price is that in the final configuration, when we trace an arbitrary component flow ${\vec v_{{A_i}{{\left\{ \gamma  \right\}}^{{\rm{cr}}}}{A_j}}}$ describing part of the entanglement between the boundary elementary regions ${A_i}$ and ${A_j}$, we find that the corresponding thread bundle only relates some special part of the two regions, which is unnatural as we mentioned earlier. From another viewpoint, the equivalence of describing the net result seems to imply the possibility that one can reorganize the information of the boundary quantum system more economically in reconstructing the bulk geometry in the holography setting.

In order to express the direct relation between the distribution of bit threads in a single elementary cell and the $W$ isometry tensor in the OSED tensor network more explicitly, it is useful to reorganize and redefine the flow in a single cell. For simplicity, taking the cell $a$ in figure~\ref{fig3  }(a) as an example, one can see that the sources of bit threads within it is more complicated than the simpler case in figure~\ref{fig1}. On surface $1$, there are thread bundle connecting elementary regions $A$ and $C$, and that connecting $A$ and the complement of $ABC$, i.e., $K$ passing through. Similarly, on surface $2$, there are threads connecting $B$ and $C$, and those connecting $B$ and $K$ passing through. While on surface $3$, all the threads passing through $1$ and $2$ are involved. Naturally, one can redefine three thread bundles in net result, or equivalently, redefine three component flows, to return to the most basic diagram figure~\ref{fig2 }. The redefinition can be implemented to arbitrary elementary cell. To see this, let us still denote the three minimal surface under consideration as $1$, $2$ and $3$ respectively for convenience, then we can explicitly define three flows $\bf{{{\vec v}_{12}}}$, $\bf{{{\vec v}_{13}}}$ and $\bf{{{\vec v}_{23}}}$ required to satisfy
\be\rho \left( \bf{{{\vec v}_{12}}} \right) = \sum\limits_{i,j} {\rho \left( {{{\vec v}_{{A_i}{\gamma _1}{\gamma _2}{A_j}}}} \right)},\\
\rho \left( \bf{{{\vec v}_{13}}} \right) = \sum\limits_{i,j} {\rho \left( {{{\vec v}_{{A_i}{\gamma _1}{\gamma _3}{A_j}}}} \right)},\\
\rho \left( \bf{{{\vec v}_{23}}} \right) = \sum\limits_{i,j} {\rho \left( {{{\vec v}_{{A_i}{\gamma _2}{\gamma _3}{A_j}}}} \right)}, \ee
where we represent the reorganized flows in boldface, and ${\vec v_{{A_i}{\gamma _a}{\gamma _b}{A_j}}}$ represents the set of component flows that crosses through surfaces ${\gamma _a}$ and ${\gamma _b}$ simultaneously. It should be noted that what we essentially redefine is the thread bundle rather than the component flow. It is the thread density of the redefined thread bundle connecting surface ${\gamma _a}$ and ${\gamma _b}$ that is required to be equal to the sum of the thread densities of the original thread bundles connecting ${\gamma _a}$ and ${\gamma _b}$ everywhere in the elementary cell. Therefore, this redefinition is explicit and unique. However, the corresponding redefined component flows are not unique, because one can always change the orientation on any subset of the vector field to obtain a new flow. In particular, we will still adopt the convention that the orientation of the redefined component flows is consistent with figure~\ref{fig2  }. This kind of redefinition within a local cell is in essence an idea of ``reduction'', i.e., instead of pursuing the detailed entanglement structure of a whole spacetime boundary, we use a reduced mixed state to effectively describe the entanglement information between a subsystem and its complement in the whole full state.

In a word, as shown in figure~\ref{fig3  }(c), we have thus obtained the bit thread representation of the fundamental tensor $W$ tensor in the OSED tensor network, which can faithfully represents the discretization of spacetime. We find that each $W$ tensor associated with a cell actually contains the entanglement information between the different pairs of elementary region in the boundary system. In terms of the reorganized flows above, the conditions that the locking multiflow within each elementary cell must satisfy can be summarized as follows,
\be\nabla  \cdot {\bf{\vec v_{ab}}} = 0,\\
\rho ({\bf V}) \le 1,\\
{\left. {\hat n \cdot \bf{{\vec v}_{ab}}} \right|_{{\gamma _c} \ne {\gamma _a}\,{\rm{or}}\,{\gamma _b}}} = 0,\\
{\bf{\vec v_{ab}}} \bot {\gamma _a}\,{\rm{and}}\,{\gamma _b},\\
{\left. {\rho \left( {\bf V} \right)} \right|_{{\gamma _a}\,{\rm{or}}\,{\gamma _b}}} = {\sum\limits_{a < b} {\left. {\left| \bf{{{\vec v}_{ab}}} \right|} \right|} _{{\gamma _a}\,{\rm{or}}\,{\gamma _b}}} = 1,\ee
where $a$ and $b$ can be respectively taken as $1$, $2$ and $3$ representing the boundaries of the elementary cell. This set of conditions can be referred to as the OSED conditions.

We have shown that, dividing the boundary quantum system into a set of adjacent non-overlapping elementary regions ${A_1}, \ldots ,{A_n}$, a set of thread bundles connecting pairs of elementary regions (which can be described by a multiflow $V$ satisfying the conditions eqs.((\ref{req1})-(\ref{req5})) is equivalent to an OSED tensor network corresponding this division. In fact, this kind of division of the boundary system can be regarded as a coarse-graining description of the entanglement structures of boundary quantum system. Therefore, our investigation leads to such a point of view, i.e., the collection of these thread bundles manifests how the boundary entanglement structures at some coarse-grained level to emerge a discretized bulk spacetime matching this coarse-grained level.

It is appropriate to make a few comments at this stage. Firstly, we emphasized that what is proposed to correspond with the OSED tensor network is the set of the thread bundles, rather than the component flows. Because the map between flows and thread configurations is not one-to-one, when locally changing the orientation on part of a flow, it still represents the same underlying thread configuration entity. However, we can always adopt a convenient convention to represent the set of thread bundles with a particular multiflow.

Another point worth pointing out is that, according to our understanding, the local parallelity of the thread bundles can be naturally understanded as a manifestation of the coarse-graining effect. As mentioned earlier, a problem about the concept of thread bundle is that, when we use a component flow ${{\vec v}_{ij}}$ to describe the thread bundle representing the entanglement between two elementary regions ${A_i}$ and ${A_j}$ mathematically, the threads in the thread bundle are required to be locally parallel, which seems to result in a special pattern of entanglement structure, i.e., the outer regions of ${A_i}$ can only be connected to the outer regions of ${A_j}$, while the inner regions of ${A_i}$ can only be connected to the inner regions of ${A_j}$, otherwise it will cause the crossing of bit threads (see figure~\ref{fig2 }). Since a bit thread is usually regarded as representing a distilled EPR pair, there is no reason to think that the entanglement between the two regions should be distilled out according to such a special pattern.
\begin{figure}[htbp]     \begin{center}
		\includegraphics[height=6cm,clip]{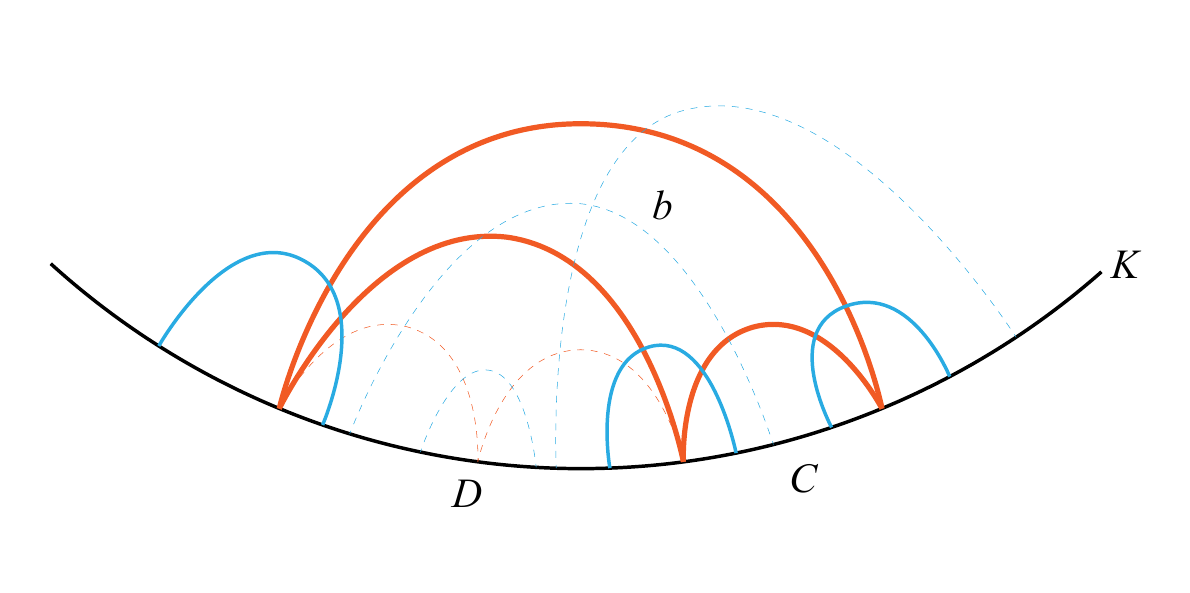}
		\caption{Removing two RT surfaces from figure~\ref{fig3  }, the number of independent locally parallel thread bundles decreases to three from six, where the red dashed lines represent the removed RT surfaces, and the blue dashed lines represent the vanishing thread bundles.}
		\label{fig5  }
	\end{center}	
\end{figure}

A natural resolution is to suggest that this locally parallelity required for the thread bundles is essentially an effect of coarse-graining. To see this, consider the case in figure~\ref{fig5  }, which is obtained by removing the RT surfaces associated with elementary regions $A$ and $B$ respectively from the figure~\ref{fig3  }(a). In order to find the thread configuration locking this bulk cell decomposition, on the one hand, we can regard the union of $A$ and $B$ as a single elementary region $D$, and then assign the thread bundle to each pair of elementary regions, as shown in figure~\ref{fig5  }. In this case, we need three independent locally parallel thread bundles in total. On the other hand, the locking thread configuration in figure~\ref{fig3  } can also satisfy the locking conditions here. In this case, however, the number of independent locally parallel thread bundles increases to six. Comparing the two thread configurations, we find that this is tantamount to replacing the set of two intersecting thread bundles $A$-$K$ and $B$-$K$ in figure~\ref{fig3  }(a) with a single locally parallel thread bundle $D$-$K$~\footnote{For convenience, we denote the thread bundle connecting $A$ and $K$ simply as $A$-$K$, etc.}. Similarly, $A$-$C$ and $B$-$C$ are replaced with $D$-$C$. Furthermore, in figure~\ref{fig3  }(a) there is an extra thread bundle $A$-$B$ representing the entanglement between the regions $A$ and $B$, i.e., the internal entanglement within region $D$. Therefore, the description of local parallelity of the thread bundle is a manifestation of coarse-graining. If one hopes to describe the geometric structure of the dual spacetime in a more refined way, the more independent thread bundles representing the entanglement structures of the boundary quantum system are needed, i.e., the quantum system is required to be divided into more elementary regions.

In general the locking thread configuration, i.e., the solution of the optimal flux problem for a certain set of elementary regions in not unique, as we have just illustrated. However, we have shown that these optimized locking thread configurations are not all equivalent to each other. Instead, some of them actually correspond to a more refined exploration of the structure of spacetime, because they can provide more detailed information about the entanglement structures of the quantum system, or in other words, they actually correspond to different OSED tensor network. It should be noted that, however, the map between the set of thread bundles satisfying the conditions~eqs.((\ref{req1})-(\ref{req5})) and the OSED tensor network is not one-to-one. As mentioned earlier, the locking thread configuration corresponding to a bulk-cell decomposition is not unique. Therefore, more precisely, the OSED tensor network is equivalent to a certain equivalent class of locking thread configuration satisfying conditions~eqs.((\ref{req1})-(\ref{req5})). It might be interesting to consider whether it can further classify the locking thread configurations that describe the more refined structure of spacetime.

\section{Bit thread and entanglement of purification}\label{sec4}
In the previous section, we only discuss the cases involving the minimal surfaces whose anchor points are located on the boundary of the whole spacetime. For more general surface growth pictures involving bulk minimal surfaces, the spirit of the discussion is the same, although the details of discussion will be more complicated. However, there is a particular case involving bulk minimal surfaces worth noting. That is, in fact the holographic entanglement of purification process can be considered as a special surface growth process, or equally, an OSED tensor network.

\subsection{A brief review of entanglement of purification}
Dividing any quantum system into two parts ${A_1}$ and ${A_2}$, there is an quantum information theoretic quantity, namely entanglement of purification (EoP) ${E_P}({A_1}:{A_2})$ to measure the correlations between ${A_1}$ and ${A_2}$~\cite{EOP}. The entanglement of purification is defined as follows: let ${\rho _{{A_1}{A_2}}}$ be a density matrix on this bipartite system ${H_{{A_1}}} \otimes {H_{{A_2}}}$. Let $\left| \psi  \right\rangle  \in {H_{{A_1}{A'_1}}} \otimes {H_{{A_2}{A'_2}}}$ be a purification of ${\rho _{{A_1}{A_2}}}$ by introducing two auxiliary systems ${A'_1}$ and ${A'_2}$, i.e.,
\be {{\rm Tr}_{{A'_1}{A'_2}}}\left| \psi  \right\rangle \left\langle \psi  \right| = {\rho _{{A_1}{A_2}}}.\ee
Then the EoP of ${\rho _{{A_1}{A_2}}}$ is defined as
\be {E_P}({A_1}:{A_2}) = \mathop {\min }\limits_{{{\left| \psi  \right\rangle }_{{A_1}{A'_1}{A_2}{A'_2}}}} S({A_1}{A'_1}),\ee
where we minimize over all possible purifications $\left| \psi  \right\rangle $ of ${\rho _{{A_1}{A_2}}}$ in all possible auxiliary Hilbert spaces ${H_{{A'_1}}}$ and ${H_{{A'_2}}}$, and $S\left( {{A_1}{A'_1}} \right)$ is the von Neumann entropy of the reduced density matrix obtained by tracing out the ${{A_2}{A'_2}}$ part of $\left| \psi  \right\rangle \left\langle \psi  \right|$.

Quite similar to the RT formula~eq.(\ref{rt}) for holographic entanglement entropy, there is also a holographic formula for EoP in the context of AdS/CFT correspondence. For two boundary subregions ${A_1}$ and ${A_2}$ of a holographic CFT state, it has been conjectured in~\cite{Takayanagi:2017knl,Nguyen:2017yqw} that ${E_P}({A_1}:{A_2})$ is given to leading order in ${G_N}$ by the area of the entanglement wedge cross section ${\Gamma _{{A_1}:{A_2}}}$. The entanglement wedge cross section ${\Gamma _{{A_1}:{A_2}}}$ is a surface of minimal area anchored to the boundary of the entanglement wedge ${W_{{A_1}{A_2}}}$ of ${A_1} \cup {A_2}$, such that ${\Gamma _{{A_1}:{A_2}}}$ partitions ${W_{{A_1}{A_2}}}$ into a region that is entirely adjacent to ${A_1}$ and another region that is entirely adjacent to ${A_2}$. In our units (where $4{G_ N} = 1$), that is
\be\label{HEOP}{E_P}({A_1}:{A_2}) = {\rm{Area}}\left( {{\Gamma _{{A_1}:{A_2}}}} \right).\ee

It is worth pointing out that, besides EoP, there are also other quantum information theoretical quantities proposed to be associated with the entanglement wedge cross section, such as the reflected entropy~\cite{Dutta:2019gen}, the logarithmic negativity~\cite{Kudler-Flam:2018qjo,Kusuki:2019zsp}, the ``odd entropy''~\cite{Tamaoka:2018ned}, the ``differential purification''~\cite{Espindola:2018ozt}, and the balanced partial entanglement~\cite{Wen:2021qgx}, etc. However, since our work is based on the tensor network description and the surface/state correspondence, we will focus only on the EoP proposal.

\subsection{The bit thread viewpoint of holographic EoP }\
There have been some work investigating the holographic entanglement of purification from the view point of bit thread~\cite{Du:2019emy,Bao:2019wcf,Harper:2019lff,Agon:2018lwq,Hubeny:2018bri,Ghodrati:2019hnn}, and the bit thread interpretation of EoP has been proposed in~\cite{Du:2019emy} (see also~\cite{Harper:2019lff,Bao:2019wcf}). However, from the relation between bit thread and entanglement distillation presented in this paper, it is natural to propose an alternative bit thread interpretation for holographic EoP. It turns out that our new interpretation is different from the existing interpretations in the previous work.
\begin{figure}[htbp]     \begin{center}
		\includegraphics[height=8cm,clip]{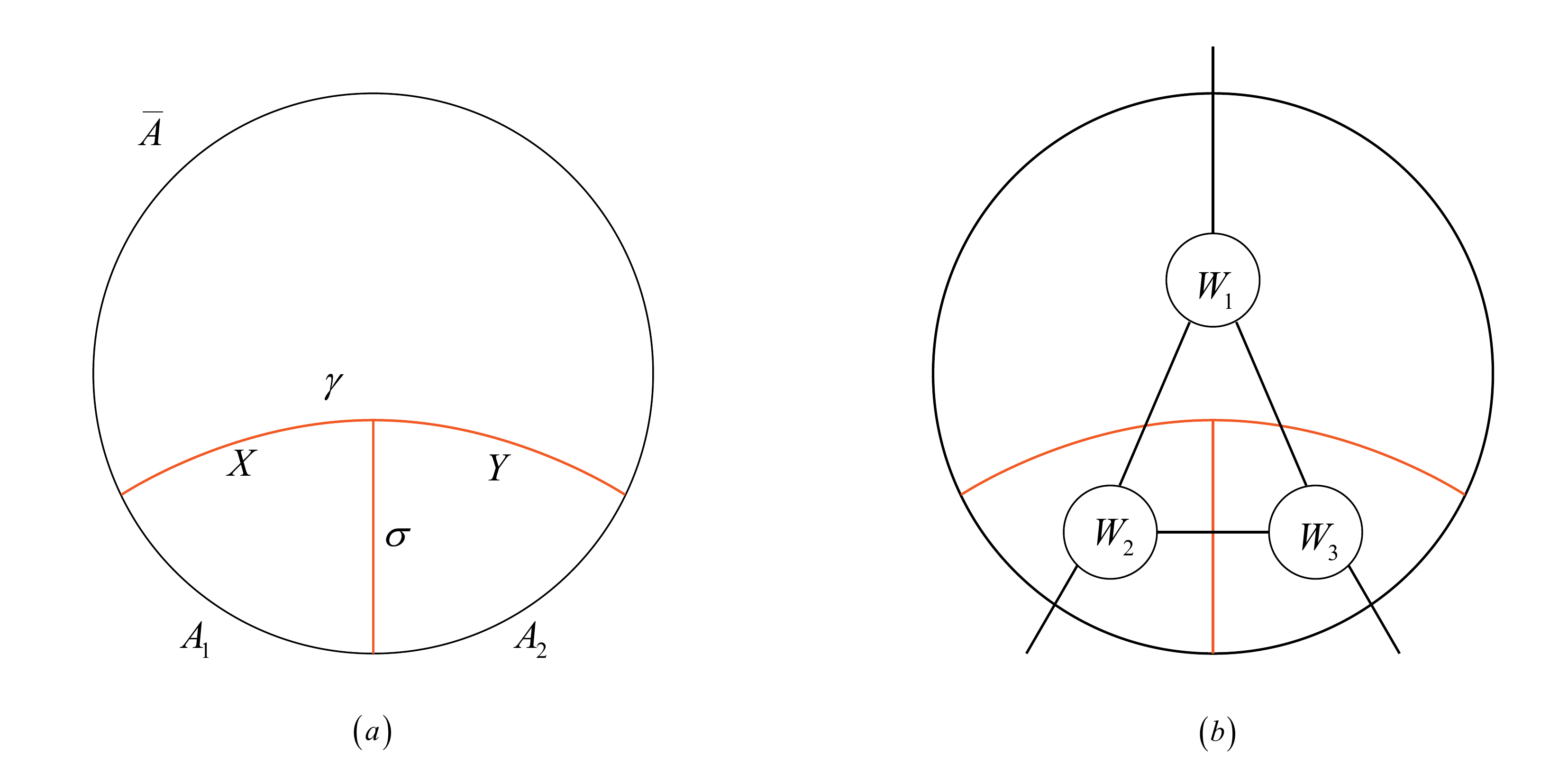}
		\caption{(a) The holographic EoP can be naturally regarded as a surface growth process. (b) The holographic EoP can correspond to an OSED tensor network.}
		\label{fig6  }
	\end{center}	
\end{figure}

Firstly, we will show that, according to the understanding of the surface growth process in~\cite{Lin:2020thc}, one can prove the holographic EoP formula~eq.(\ref{HEOP}). In other words, holographic EoP can be naturally regarded as a surface growth process, i.e., it can correspond to an OSED tensor network~\footnote{Actually, \cite{Bao:2018pvs} also proposed a kind of OSED tensor network corresponding to holographic EoP, but their idea is to assume the holographic EoP is correct in advance and then use it to construct the OSED tensor network. While what we utilize is the idea in~\cite{Lin:2020thc}, which admitted the surface/state correspondence in advance and then constructed the more general OSED tensor networks. More precisely, we will prove the holographic HoP is a special case of the latter.}. As shown in figure~\ref{fig6  }(a), we want to calculate the EoP of the bipartite region ${A_1}{A_2}$ in the holographic framework. For this purpose, we first intuitively image that a minimal surface $\gamma $ grows from ${\bar A}$, i.e., the complement of ${A_1}{A_2}$. Then, taking an anchor point on surface $\gamma $, which divides $\gamma $ into two parts, $X$ and $Y$, at the same time, taking the segmentation point of ${A_1}$ and ${A_2}$ as another anchor point, by surface growth process, a new minimal surface $\sigma $ can continue to grow. Now, according to the ``surface growth rule'' proposed in~\cite{Lin:2020thc}, i.e., the generalized RT formula, one obtain
\be\label{grt} {\rm Area}(\sigma ) = S({X{A_1}}),\ee
where $S({X{A_1}})$ denotes the von Neumann entropy of the union of $X$ and ${A_1}$. As also indicated in~\cite{Du:2019emy} explicitly (and implicitly utilized in other related papers), due to the surface/state correspondence, the closed surface ${A_1}{A_2}XY$ corresponds to a pure state $\left| {{A_1}{A_2}XY} \right\rangle $, which realizes a purificaiton for ${\rho _{{A_1}{A_2}}}$. Furthermore, in the framework of the surface/state correspondence, the minimal surface $XY$ has no internal entanglement, and thus has the minimal possible Hilbert space dimension to purify ${\rho _{{A_1}{A_2}}}$. Then by identifying
\be X &=& {A'_1},\\
Y &=& {A'_2},\ee
we have immediately
\be {E_P}({A_1}:{A_2}) = \mathop {\min }\limits_{{{\left| \psi  \right\rangle }_{{A_1}{A'_1}{A_2}{A'_2}}}} S({A_1}{A'_1}) = \min S(X{A_1}),\ee
by eq.(\ref{grt}), one obtains
\be {E_P}({A_1}:{A_2}) = \min {\rm{Area}}(\sigma ) \equiv {\rm{Area}}\left( {{\Gamma _{{A_1}:{A_2}}}} \right),\ee
which is exactly the holographic EoP proposal~eq.(\ref{HEOP}).

\begin{figure}[htbp]     \begin{center}
		\includegraphics[height=8cm,clip]{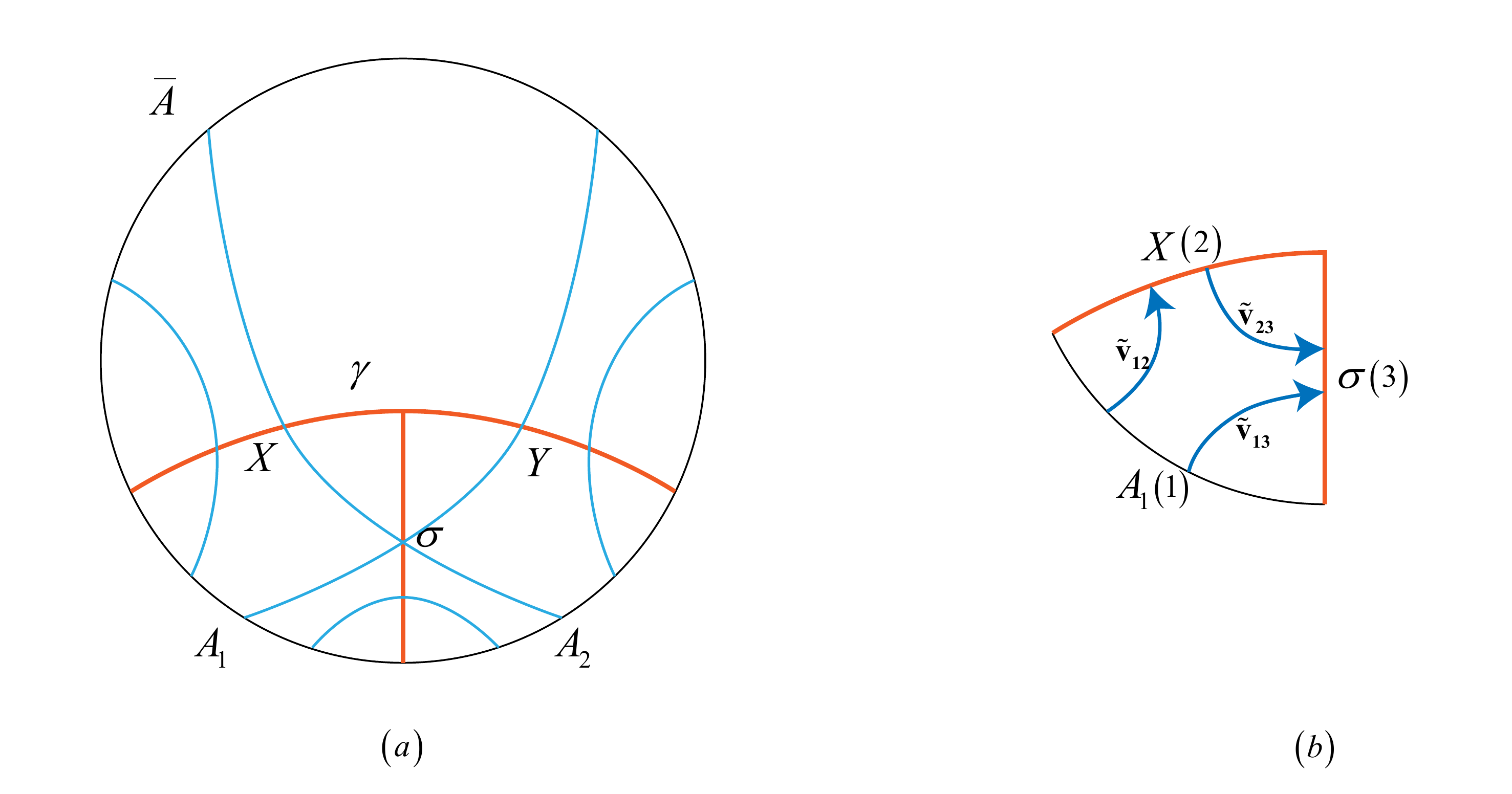}
		\caption{(a) The bit thread interpretation for holographic EoP. (b) Recombining the flows into three redefined flows in the cell.}
		\label{fig7  }
	\end{center}	
\end{figure}

Actually, as mentioned above, the idea of utilizing the surface/state correspondence and regarding the closed surface ${A_1}{A_2}XY$ as a pure state has been indicated or implied in the previous related work. However, the novel point of our reinvestigation is that, from the viewpoint that the surface growth process should be understood as a kind of entanglement distillation and correspond to an OSED tensor network as shown in figure~\ref{fig6  }(b), the entropy on surface $\sigma $ visibly results from the disentanglement effect between ${A_1}$ and bulk minimal surface $X$. Therefore, in our framework of surface growth scheme and OSED tensor network, one can naturally obtain a bit thread interpretation for holographic EoP process as shown in figure~\ref{fig7  }(a). More specifically, there are four independent thread bundles involved within the cell bounded by surfaces ${A_1}$, $X$ and $\sigma $, which are represented by four component flows ${\vec v_{\bar AX{A_1}}}$, ${{\vec v}_{\bar AX\sigma {A_2}}}$, ${{\vec v}_{\bar AY\sigma {A_1}}}$ and ${{\vec v}_{{A_1}\sigma {A_2}}}$ respectively. As shown in figure~\ref{fig7  }(b), we can also recombine these flows into three redefined flows satisfying
\be\rho \left( {\bf{{\vec v}_{12}}} \right) &=& \rho \left( {{{\vec v}_{\bar AX{A_1}}}} \right),\\
\rho \left( {\bf{{\vec v}_{13}}} \right) &=& \rho \left( {{{\vec v}_{\bar AY\sigma {A_1}}}} \right) + \rho \left( {{{\vec v}_{{A_1}\sigma {A_2}}}} \right),\\
\rho \left( {\bf{{\vec v}_{23}}} \right) &=& \rho \left( {{{\vec v}_{\bar AX\sigma {A_2}}}} \right),\ee
where for simplicity and consistency with the previous section, we denote surface ${A_1}$ as $1$, $X$ as $2$, and $\sigma$ as $3$ respectively.
\begin{figure}[htbp]     \begin{center}
		\includegraphics[height=13cm,clip]{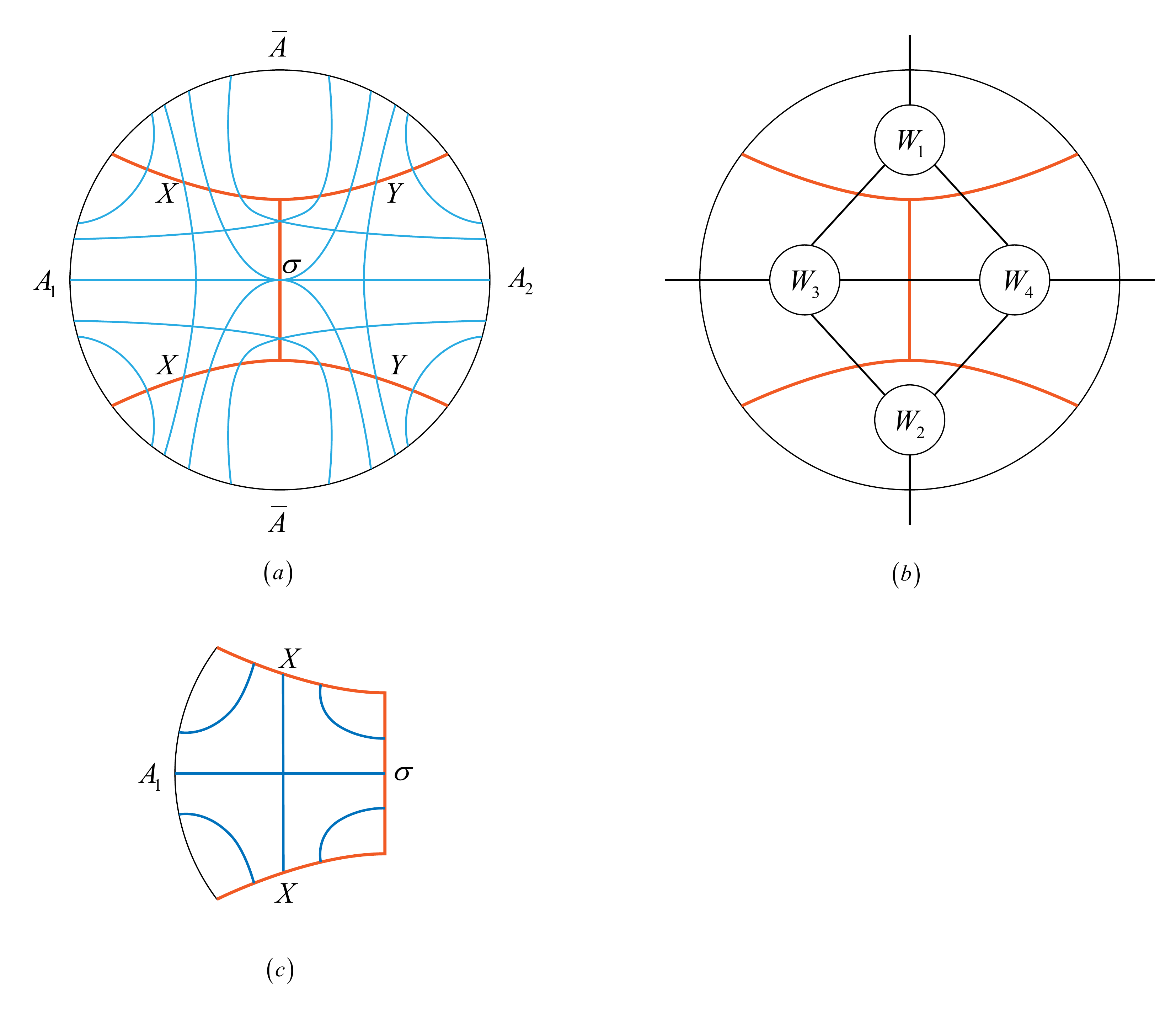}
		\caption{(a) The bit thread interpretation of holographic EoP for the general case. (b) The corresponding OSED tensor network in this case. (c) Redefining the reorganized flows in a cell with four boundaries.}
		\label{fig8  }
	\end{center}	
\end{figure}

One can note that, there is an apparent difference between our bit thread interpretation of holographic EoP and the existing interpretation in~\cite{Du:2019emy} (see also~\cite{Harper:2019lff,Bao:2019wcf}). The existing bit thread interpretation for holographic EoP can be briefly described as follows. In their framework, the maximal flux of bit threads starting from ${A_1}$ contribute two parts. One part fully provides the maximal thread flux on surface $\sigma $, while the rest part runs into surface $X$. Therefore, in their scenario, there are no bit threads connecting $X$ and $\sigma $. While in our framework, the threads connecting $X$ and $\sigma $, i.e., the thread bundle represented by ${\bf{{\vec v}_{23}}} $ is also necessary in order for the locking thread configuration to faithfully reflect this process. The difference between these two framework can also be rephrased in another way. In the former framework, the generalization of Riemannian max flow-min cut (MFMC) theorem was used, and the bit thread flux are thus not required to be maximal on the $X$ surface, or in our language, the bit thread are not required to lock $X$. While in our framework, since we identify the locking thread configuration constructed by the bulk cell decomposition with an OSED tensor network, the bit thread should maximize the thread flux in the $X$ surface (regardless of whether or not there is a ${\bf{{\vec v}_{23}}} $ flow passing through $X$).

It is worthwhile to further clarify that our requirements are reasonable and natural in our framework. Firstly, according to the understanding of the (generalized) OSED tensor network, all the bulk minimal surfaces should play the same role, therefore there should be no particular reason to think that the bit threads should saturate the density bound on the minimal surfaces like $\sigma $, while do not saturate the density bound on the minimal surfaces like $X$. More importantly, there is also a physical reason. We have emphasized in the previous that in each step of ``surface growth process'', in order for the state of the union of the two previous surfaces being mapped to the state of the new minimal surface by an isometry tensor, there must exist necessary entanglement between the two previous surfaces, which can be naturally represented by the bit threads connecting them. Based on this, one can see that ${\bf{{\vec v}_{23}}} $ is necessary in our framework. The reason is as follows: as shown in figure~\ref{fig6 }, except that the state on the union of $X$ and ${A_1}$ should be mapped into the state on  $\sigma $ by the isometry tensor, the state on the union of $\sigma $ and $Y$ should also be mapped into that of ${A_2}$ by the isometry tensor. Therefore, in this latter case, there must be entanglement between $\sigma $ and $Y$, and the bit threads describing this entanglement cannot be ignored. From the symmetry, there must also be entanglement between $\sigma $ and $X$, which means that the ${\bf{{\vec v}_{23}}} $ component is necessary in our framework. Or in an equivalent perspective, imaging the ``surface growth process'' in an opposite way: first ${\bar A}$ ``generates'' $\gamma$, then, the union of $Y$ and ${A_2}$ ``generates'' $\sigma $, therefore, at the final, the state of the union of $\sigma $ and $X$ should be mapped into the state of ${A_1}$ by an isometry tensor, thus the bit threads connecting $\sigma $ and $X$ represented by${\bf{{\vec v}_{23}}} $ is necessary.

The situation will be more apparent by expressing the above discussions in mathematical form. In our framework, if we choose ${\bf{{\vec v}_{23}}} = 0$, then, we have
\be S(1)&=& N(1),\\
S(3)&=& N(3),\ee
as in~eq.(\ref{s3}) and eq.(\ref{s1}). Recalling that in a locking thread configuration corresponding to an OSED tensor network, similar to eq.(\ref{f112}) we have
\be S\left( 2 \right) + S\left( 3 \right) - S\left( 1 \right) = 2F{\left( 2 \right)_{23}},\ee
in which
\be F{(2)_{23}} = \left| {\int_{{\gamma _2}} {\bf{{\vec v}_{23}}} } \right|.\ee
Supposing ${\bf{{\vec v}_{23}}} = 0$, one would obtain
\be\label{inc} S\left( 2 \right) + S\left( 3 \right) - S\left( 1 \right) = 0,\ee
which is, however, inconsistent with an obvious fact. According to the surface/state correspondence, the von Neumann entropy of the minimal surface $X$ or $\sigma $ (or in the language of bit thread, the locking thread flux through it) is equal to its area per se, i.e.,
\be{\rm{Area}}\left( X \right) &=& S\left( 2 \right),\\
{\rm{Area}}\left( \sigma  \right) &=& S\left( 3 \right).\ee
On the other hand, the von Neumann entropy of ${A_1}$ region should be equal to the area of the minimal surface ${\gamma \left( {{A_1}} \right)}$ homologous to it,
\be{\rm{Area}}\left( {\gamma \left( {{A_1}} \right)} \right) = S\left( 1 \right).\ee
Since surface $X \cup \sigma $ is also homologous to ${A_1}$, the sum of the area of surface $X$ and that of $\sigma $ must be greater than the area of the minimal surface homologous to ${A_1}$, we must have
\be{\rm{Area}}\left( X \right) + {\rm{Area}}\left( \sigma  \right) - {\rm{Area}}\left( {\gamma \left( {{A_1}} \right)} \right) > 0,\ee
which leads to
\be S\left( 2 \right) + S\left( 3 \right) - S\left( 1 \right) > 0,\ee
and is inconsistent with eq.(\ref{inc}). 
Therefore, in our framework that the locking thread configuration is identified with a corresponding OSED tensor network, the ${\bf{{\vec v}_{23}}} $ component is inevitable, we thus obtain a different but natural bit thread interpretation of holographic EoP. In a word, based on the viewpoint that the locking thread configuration is equivalent to an OSED tensor network, we show that the bit thread interpretation of holographic EoP is described by the locking thread configuration as shown in figure~\ref{fig7 }(a). The advantage of our interpretation version is that it is obtained in a more general framework, and naturally manifests all the necessary physical entanglements involved in the process. 

A similar investigation is also applicable for the general case that the two boundary subregions ${A_1}$ and ${A_2}$ is not adjacent, except that the threads to be traced are more complicated, see figure~\ref{fig8  }. Similarly, one can also redefine the reorganized flows in such a cell with four boundaries, as shown in figure~\ref{fig8  }(c).

\section{Conclusions and discussions}\label{sec5}
In this paper, we provide a bit thread interpretation for the OSED tensor network (which is also named as ``surface growth scheme'' intuitively) proposed in~\cite{Bao:2018pvs} and later generalized in~\cite{Lin:2020thc}. More specifically, based on the locking theorem of bit thread proved by the bulk-cell gluing method in~\cite{Headrick:2020gyq}, we match a class of locking thread configurations satisfying a set of conditions with the OSED tensor network, and argue that this class of locking thread configurations describe the entanglement details of the OSED tensor network. In this way, we show the connection between bit thread and entanglement distillation and obtain the explicit bit thread interpretation of the $W$ tensor in the OSED tensor network. On the one hand, the locking bit thread configuration can provide a detailed description for the discretization of spacetime. On the other hand, the OSED tensor network provides a picture of reconstructing the spacetime with the surface growth scheme, i.e., the emergence of spacetime can be regarded as the reorganization of the boundary degree of freedom through the entanglement distillation. Therefore, the bit thread perspective will provide some new insights into the bulk reconstruction scheme in the framework of the holographic principle.

In particular, we find that in the framework of surface/state correspondence, the holographic EoP process can be regarded as a special case of surface growth process. Based on this viewpoint, we naturally obtain a bit thread interpretation for holographic EoP, which is a class of locking thread configurations corresponding to the OSED tensor network associated with this process. It turns out that our interpretation for holographic EoP must be different from the existing interpretations. However, the virtue of our interpretation version is that it is obtained in a more general framework, thus it is more natural in physics. Moreover, in our framework, our interpretation manifest all the necessary physical entanglements involved in the process, which is more physical in a sense.

\section*{Acknowledgement}
We would like to thank D.-H. Du, C.-B. Chen and F.-W. Shu for helpful discussions. J.R.S. was supported by the National Natural Science Foundation of China (No.~11675272). The Project is also funded by China Postdoctoral Science Foundation
(No. 2019M653137)






\begin{references}
	
	
	\bibitem{Maldacena:1997re}
	J.~M.~Maldacena,
	``The Large N limit of superconformal field theories and supergravity,''
	Int. J. Theor. Phys. \textbf{38}, 1113-1133 (1999)
	[arXiv:hep-th/9711200 [hep-th]].
	
	\bibitem{Gubser:1998bc}
	S.~S.~Gubser, I.~R.~Klebanov and A.~M.~Polyakov,
	``Gauge theory correlators from noncritical string theory,''
	Phys. Lett. B \textbf{428}, 105-114 (1998)
	[arXiv:hep-th/9802109 [hep-th]].
	
	\bibitem{Witten:1998qj}
	E.~Witten,
	``Anti-de Sitter space and holography,''
	Adv. Theor. Math. Phys. \textbf{2}, 253-291 (1998)
	[arXiv:hep-th/9802150 [hep-th]].
	


    \bibitem{VanRaamsdonk:2010pw}
    M.~Van Raamsdonk,
    ``Building up spacetime with quantum entanglement,''
    Gen. Rel. Grav. \textbf{42}, 2323-2329 (2010)
    [arXiv:1005.3035 [hep-th]].

    \bibitem{Lashkari:2013koa}
    N.~Lashkari, M.~B.~McDermott and M.~Van Raamsdonk,
    ``Gravitational dynamics from entanglement 'thermodynamics',''
    JHEP \textbf{04}, 195 (2014)
    [arXiv:1308.3716 [hep-th]].

    \bibitem{Faulkner:2013ica}
    T.~Faulkner, M.~Guica, T.~Hartman, R.~C.~Myers and M.~Van Raamsdonk,
    ``Gravitation from Entanglement in Holographic CFTs,''
    JHEP \textbf{03}, 051 (2014)
    [arXiv:1312.7856 [hep-th]].

    \bibitem{Faulkner:2017tkh}
    T.~Faulkner, F.~M.~Haehl, E.~Hijano, O.~Parrikar, C.~Rabideau and M.~Van Raamsdonk,
    ``Nonlinear Gravity from Entanglement in Conformal Field Theories,''
    JHEP \textbf{08}, 057 (2017)
    [arXiv:1705.03026 [hep-th]].

    \bibitem{Agon:2020mvu}
    C.~A.~Ag\'on, E.~C\'aceres and J.~F.~Pedraza,
    ``Bit threads, Einstein's equations and bulk locality,''
    [arXiv:2007.07907 [hep-th]].

    \bibitem{Bao:2019bib}
    N.~Bao, C.~Cao, S.~Fischetti and C.~Keeler,
    ``Towards Bulk Metric Reconstruction from Extremal Area Variations,''
    Class. Quant. Grav. \textbf{36}, no.18, 185002 (2019)
    [arXiv:1904.04834 [hep-th]].

    \bibitem{Sun:2019ycv}
    J.~R.~Sun and Y.~Sun,
    ``On the emergence of gravitational dynamics from tensor networks,''
    [arXiv:1912.02070 [hep-th]].
	
	\bibitem{Ryu:2006bv}
	S.~Ryu and T.~Takayanagi,
	``Holographic derivation of entanglement entropy from AdS/CFT,''
	Phys. Rev. Lett. \textbf{96}, 181602 (2006)
	[arXiv:hep-th/0603001 [hep-th]].
	
	\bibitem{Ryu:2006ef}
	S.~Ryu and T.~Takayanagi,
	``Aspects of Holographic Entanglement Entropy,''
	JHEP \textbf{08}, 045 (2006)
	[arXiv:hep-th/0605073 [hep-th]].
	
	\bibitem{Hubeny:2007xt}
	V.~E.~Hubeny, M.~Rangamani and T.~Takayanagi,
	``A Covariant holographic entanglement entropy proposal,''
	JHEP \textbf{07}, 062 (2007)
	[arXiv:0705.0016 [hep-th]].


	
	
	\bibitem{Bao:2018pvs}
	N.~Bao, G.~Penington, J.~Sorce and A.~C.~Wall,
	``Beyond Toy Models: Distilling Tensor Networks in Full AdS/CFT,''
	JHEP \textbf{19}, 069 (2020)
	[arXiv:1812.01171 [hep-th]].
	
	
	\bibitem{Lin:2020thc}
	Y.~Y.~Lin, J.~R.~Sun and Y.~Sun,
	``Surface growth scheme for bulk reconstruction and tensor network,''
	JHEP \textbf{12}, 083 (2020)
	[arXiv:2010.03167 [hep-th]].

	
	\bibitem{Miyaji:2015yva}
	M.~Miyaji and T.~Takayanagi,
	``Surface/State Correspondence as a Generalized Holography,''
	PTEP \textbf{2015}, no.7, 073B03 (2015)
	[arXiv:1503.03542 [hep-th]].
	
	\bibitem{Miyaji:2015fia}
	M.~Miyaji, T.~Numasawa, N.~Shiba, T.~Takayanagi and K.~Watanabe,
	``Continuous Multiscale Entanglement Renormalization Ansatz as Holographic Surface-State Correspondence,''
	Phys. Rev. Lett. \textbf{115}, no.17, 171602 (2015)
	[arXiv:1506.01353 [hep-th]].


	
	

    \bibitem{Freedman:2016zud}
    M.~Freedman and M.~Headrick,
    ``Bit threads and holographic entanglement,''
    Commun. Math. Phys. \textbf{352}, no.1, 407-438 (2017)
    [arXiv:1604.00354 [hep-th]].

    \bibitem{Cui:2018dyq}
    S.~X.~Cui, P.~Hayden, T.~He, M.~Headrick, B.~Stoica and M.~Walter,
    ``Bit Threads and Holographic Monogamy,''
    Commun. Math. Phys. \textbf{376}, no.1, 609-648 (2019)
    [arXiv:1808.05234 [hep-th]].

    \bibitem{Headrick:2020gyq}
    M.~Headrick, J.~Held and J.~Herman,
    ``Crossing versus locking: Bit threads and continuum multiflows,''
    [arXiv:2008.03197 [hep-th]].
    
    \bibitem{Bao:2020uku}
    N.~Bao and J.~Harper,
    ``Bit threads on hypergraphs,''
    [arXiv:2012.07872 [hep-th]].


    \bibitem{EOP}
    B.M. Terhal, M. Horodecki, D.W. Leung and D.P. DiVincenzo,
    ``The entanglement of purification,''
    J. Math. Phys. 43 (2002) 4286
    [quant-ph/0202044].

    \bibitem{Takayanagi:2017knl}
    T.~Takayanagi and K.~Umemoto,
    ``Entanglement of purification through holographic duality,''
    Nature Phys. \textbf{14}, no.6, 573-577 (2018)
    [arXiv:1708.09393 [hep-th]].

    \bibitem{Nguyen:2017yqw}
    P.~Nguyen, T.~Devakul, M.~G.~Halbasch, M.~P.~Zaletel and B.~Swingle,
    ``Entanglement of purification: from spin chains to holography,''
    JHEP \textbf{01}, 098 (2018)
    [arXiv:1709.07424 [hep-th]].
    
    \bibitem{Dutta:2019gen}
    S.~Dutta and T.~Faulkner,
    ``A canonical purification for the entanglement wedge cross-section,''
    JHEP \textbf{03}, 178 (2021)
    [arXiv:1905.00577 [hep-th]].
    
    \bibitem{Kudler-Flam:2018qjo}
    J.~Kudler-Flam and S.~Ryu,
    ``Entanglement negativity and minimal entanglement wedge cross sections in holographic theories,''
    Phys. Rev. D \textbf{99}, no.10, 106014 (2019)
    [arXiv:1808.00446 [hep-th]].
    
    \bibitem{Kusuki:2019zsp}
    Y.~Kusuki, J.~Kudler-Flam and S.~Ryu,
    ``Derivation of Holographic Negativity in AdS$_3$/CFT$_2$,''
    Phys. Rev. Lett. \textbf{123}, no.13, 131603 (2019)
    [arXiv:1907.07824 [hep-th]].
    
    \bibitem{Tamaoka:2018ned}
    K.~Tamaoka,
    ``Entanglement Wedge Cross Section from the Dual Density Matrix,''
    Phys. Rev. Lett. \textbf{122}, no.14, 141601 (2019)
    [arXiv:1809.09109 [hep-th]].
    
    \bibitem{Espindola:2018ozt}
    R.~Esp\'\i{}ndola, A.~Guijosa and J.~F.~Pedraza,
    ``Entanglement Wedge Reconstruction and Entanglement of Purification,''
    Eur. Phys. J. C \textbf{78}, no.8, 646 (2018)
    [arXiv:1804.05855 [hep-th]].
    
    \bibitem{Wen:2021qgx}
    Q.~Wen,
    ``Balanced Partial Entanglement and the Entanglement Wedge Cross Section,''
    [arXiv:2103.00415 [hep-th]].


    \bibitem{Du:2019emy}
    D.~H.~Du, C.~B.~Chen and F.~W.~Shu,
    ``Bit threads and holographic entanglement of purification,''
    JHEP \textbf{08}, 140 (2019)
    [arXiv:1904.06871 [hep-th]].

    \bibitem{Bao:2019wcf}
    N.~Bao, A.~Chatwin-Davies, J.~Pollack and G.~N.~Remmen,
    ``Towards a Bit Threads Derivation of Holographic Entanglement of Purification,''
    JHEP \textbf{07}, 152 (2019)
    [arXiv:1905.04317 [hep-th]].

    \bibitem{Harper:2019lff}
    J.~Harper and M.~Headrick,
    ``Bit threads and holographic entanglement of purification,''
    JHEP \textbf{08}, 101 (2019)
    [arXiv:1906.05970 [hep-th]].

    \bibitem{Agon:2018lwq}
    C.~A.~Ag\'on, J.~De Boer and J.~F.~Pedraza,
    ``Geometric Aspects of Holographic Bit Threads,''
    JHEP \textbf{05}, 075 (2019)
    [arXiv:1811.08879 [hep-th]].

    \bibitem{Hubeny:2018bri}
    V.~E.~Hubeny,
    ``Bulk locality and cooperative flows,''
    JHEP \textbf{12}, 068 (2018)
    [arXiv:1808.05313 [hep-th]].
    
    \bibitem{Ghodrati:2019hnn}
    M.~Ghodrati, X.~M.~Kuang, B.~Wang, C.~Y.~Zhang and Y.~T.~Zhou,
    ``The connection between holographic entanglement and complexity of purification,''
    JHEP \textbf{09}, 009 (2019)
    doi:10.1007/JHEP09(2019)009
    [arXiv:1902.02475 [hep-th]].
	
	

		
	
		
	\bibitem{Swingle:2009bg}
	B.~Swingle,
	``Entanglement Renormalization and Holography,''
	Phys. Rev. D \textbf{86}, 065007 (2012)
	[arXiv:0905.1317 [cond-mat.str-el]].
		
	\bibitem{Swingle:2012wq}
	B.~Swingle,
	``Constructing holographic spacetimes using entanglement renormalization,''
	[arXiv:1209.3304 [hep-th]].
	
	
	\bibitem{Milsted:2018san}
	A.~Milsted and G.~Vidal,
	``Geometric interpretation of the multi-scale entanglement renormalization ansatz,''
	[arXiv:1812.00529 [hep-th]].
		
	
	\bibitem{Pastawski:2015qua}
	F.~Pastawski, B.~Yoshida, D.~Harlow and J.~Preskill,
	``Holographic quantum error-correcting codes: Toy models for the bulk/boundary correspondence,''
	JHEP \textbf{06}, 149 (2015)
	[arXiv:1503.06237 [hep-th]].
	
	\bibitem{Hayden:2016cfa}
	P.~Hayden, S.~Nezami, X.~L.~Qi, N.~Thomas, M.~Walter and Z.~Yang,
	``Holographic duality from random tensor networks,''
	JHEP \textbf{11}, 009 (2016)
	[arXiv:1601.01694 [hep-th]].
	
	\bibitem{Qi:2013caa}
	X.~L.~Qi,
	``Exact holographic mapping and emergent space-time geometry,''
	[arXiv:1309.6282 [hep-th]].
	

	

	

	
	

	

	
\end{references}
\end{document}